\documentclass[aip,jcp,amsmath,amssymb,reprint,
longbibliography,groupedaddress]{revtex4-2}

\usepackage{graphicx} % \includegraphics
\usepackage{mathtools}
\usepackage{bm}  % bold Greek symbols
\usepackage{booktabs} % \toprule \midrule \bottomrule
\usepackage{multirow}
\usepackage{threeparttable} % \tnote tablenotes
\usepackage{hyperref} % hyperlinks
\usepackage{xparse} % advanced macro defintions via \NewDocumentCommand
\usepackage{siunitx} % unit typesetting
\DeclareSIUnit\hartree{\ensuremath{\mathit{E}}_h}
\usepackage{dcolumn} % D
\newcolumntype{d}[1]{D{.}{.}{#1}}
\usepackage{braket} % provides bra, ket, braket, Bra, Ket, Braket

\usepackage[version=4]{mhchem}
\usepackage{color}
\usepackage{mleftright}
\usepackage{easyReview}

\newcommand{\eu}{\mathrm{e}^}
\newcommand{\rmd}{\mathrm{d}}
\newcommand{\cm}{\ensuremath{\mathrm{cm}^{-1}}}

\newcommand{\en}{\ensuremath{\varepsilon}}
\renewcommand{\nat}{\ensuremath{n_\mathrm{at}}}

\newcommand{\rotprojH}{\ensuremath{\widetilde{H}}}
\newcommand{\eckR}{\ensuremath{\widetilde{\vec{r}}}}
\newcommand{\eckRot}{\ensuremath{\widetilde{\bm{\Omega}}}}
\newcommand{\eckmom}{\ensuremath{\Theta}}
\newcommand{\Zratio}{\ensuremath{\Phi}}

\NewDocumentCommand{\funOpN}{ o m }{ 
	\IfNoValueTF{#1}
		{\ensuremath{#2_{N}}}
		{\ensuremath{#2^{#1}_{\mkern-1mu N}}}
}

\newcommand{\app}[1]{Appendix~\ref{app:#1}}
\newcommand{\eq}[1]{Eq.~\eqref{#1}}

\newcommand{\Eqn}[1]{Equation~\eqref{#1}}
\newcommand{\eqn}[1]{Eq.~\eqref{#1}}
\newcommand{\eqs}[2]{Eqs.~\eqref{#1} and \eqref{#2}}
\newcommand{\eqt}[2]{Eqs.~\eqref{#1}--\eqref{#2}}
\newcommand{\tref}[1]{Table~\ref{tab:#1}\@}

\newcommand{\fig}[2][]{Fig.~\ref{fig:#2}#1}

\providecommand{\sref}[1]{Sec.~\ref{sec:#1}\@}
\providecommand{\Sref}[1]{Section~\ref{sec:#1}\@}
\providecommand{\srefs}[2]{Secs.~\ref{sec:#1} and \ref{sec:#2}\@}

\newcommand{\op}[1]{\ensuremath{\hat{#1}}}
\newcommand{\Hop}{\op{H}}

\DeclareMathOperator{\Tr}{Tr}

\renewcommand{\vec}[1]{\ensuremath{\bm{\mathrm{#1}}}}
\newcommand{\cvec}[1]{\bm{#1}}
\newcommand{\mat}[1]{\vec{#1}}

\ProvideDocumentCommand{\der}{ o m m }{
	\IfNoValueTF{#1}
		{\ensuremath{\frac{\rmd #2}{\rmd #3}}}
		{\ensuremath{\frac{\rmd^{#1} #2}{\rmd #3^{#1}}}}
}

\ProvideDocumentCommand{\tder}{ o m m }{ 
	\IfNoValueTF{#1}
		{\ensuremath{\rmd #2 / \rmd #3}}
		{\ensuremath{\rmd^{#1} #2 / \rmd #3^{#1}}}
}

\ProvideDocumentCommand{\pder}{ o m m }{ 
	\IfNoValueTF{#1}
		{\ensuremath{\frac{\partial #2}{\partial #3}}}
		{\ensuremath{\frac{\partial^{#1} #2}{\partial #3^{#1}}}}
}

\ProvideDocumentCommand{\tpder}{ o m m }{ 
	\IfNoValueTF{#1}
		{\ensuremath{\partial #2 / \partial #3}}
		{\ensuremath{\partial^{#1} #2 / \partial #3^{#1}}}
}

\ProvideDocumentCommand{\Therm}{ o m }{
	\IfNoValueTF{#1}
		{\ensuremath{\mleft\langle #2 \mright\rangle}}
		{\ensuremath{\mleft\langle #2 \mright\rangle}_{#1}}
}

\begin{document}

\title{Exact tunneling splittings from symmetrized path integrals}
\author{George Trenins}
\email{georgijs.trenins@phys.chem.ethz.ch}
\author{Lars Meuser}
\author{Hannah Bertschi}
\author{Odysseas Vavourakis}
\author{Reto Fl\"{u}tsch}
\author{Jeremy O. Richardson}
\email{jeremy.richardson@phys.chem.ethz.ch}
%\affiliation{Laboratory of Physical Chemistry, ETH Z\"urich, 8093 Z\"urich, Switzerland}
\affiliation{Department of Chemistry and Applied Biosciences, ETH Z\"urich, 8093 Z\"urich, Switzerland}
\date{\today}

\begin{abstract}
	We develop a new simulation technique based on path-integral 
	molecular dynamics for calculating ground-state
	tunneling splitting patterns from ratios of symmetrized
	partition functions. In particular, molecular systems are
	rigorously projected onto their  $J = 0$ rotational state
	by an ``Eckart spring'' that connects two adjacent beads in a
	ring polymer.
	Using this procedure, the tunneling splitting can be 
	obtained from thermodynamic integration at just one (sufficiently 
	low) temperature.
	Converged results are formally identical to the values 
	that would have been obtained by solving the full
	rovibrational Schr\"odinger equation on a given Born--Oppenheimer 
	potential energy surface.
	The new approach is showcased
	with simulations of hydronium and methanol, which are in 
	good agreement with wavefunction-based calculations and 
	experimental measurements. The method will be of particular use
	for the study of low-barrier methyl rotations and other floppy modes,
	where instanton theory is not valid.
\end{abstract}

\maketitle

\section{Introduction}

Quantum tunneling leads to a splitting of molecular energy 
levels that is very sensitive to the height and shape of the underlying
potential energy barrier. This makes the tunneling splitting an 
effective experimental observable for probing the non-equilibrium parts of 
a potential energy surface 
(PES).\cite{Benderskii,Keutsch2001water,
StoneBook,Khazaei2016Methyl} 
It can help uncover structural
and energetic details in molecular solids, where splittings
are typically measured by inelastic neutron 
scattering\cite{Prager1997rottunnel,Colmenero2005Neutron} or
nuclear magnetic 
resonance,\cite{Horsewill1999MeTunnelNMR,Simenas2020tunnel}
and in the gas phase, where it is extracted from 
high-resolution 
microwave\cite{Fraser1989dimer,Xu1995aMeOH,Baba1999malonaldehyde,hexamerprism}
 and 
infrared\cite{Liu1985H3Osplit,Verhoeve1989H3Osplit,Pugliano1992trimer,
Xu1997MeOHsplit,Birer2009review,Keutsch2001water} 
spectra. 

These experimental measurements are particularly effective when coupled
with high-accuracy calculations, which facilitate the 
interpretation of spectroscopic data and can be used to test the 
accuracy of potential 
energy surfaces. Some of the theoretical frameworks used to this
end involve solving the Schr\"{o}dinger 
equation,\cite{Bowman1986VSCF,Bowman2007methanol,Csaszar2012FourthAge,Bowman2016Hydronium,
Coutinho2004malonaldehyde,Yurchenko2020Hydronium,Carrington2017quantum} 
resulting in
a comprehensive set of the system's energy levels.
They must, however, rely on carefully tailored representations of the 
underlying wavefunctions, PESs, and operators in order to 
combat the exponential scaling of computational costs with system size
(\emph{curse of 
dimensionality}).\cite{Thomas2018MoreThanDozen,
Wang2003MLMCTDH,Lauvergnat2014Methanol,Carter1998Multimode}
One approach to overcome this difficulty
is diffusion Monte Carlo (DMC)---a grid-free method that samples 
the ground-state wavefunction 
stochastically.\cite{Coker1987RWK,Suhm1991DMC,Gregory1995dmc,
Quack1995DMC,Wang2008malonaldehydePES} It readily yields zero-point 
energies but
is challenging to use for tunneling splitting calculations, which are
especially difficult to converge if the splitting is small. Furthermore, such calculations may contain an uncontrolled
error, since sampling excited states requires 
specifying a nodal surface, which is not known 
a~priori.

Another class of readily scalable methods is based on the 
path-integral formulation of quantum mechanics.\cite{Feynman} 
In such approaches,
quantum-mechanical properties are recast as an integral over the configuration
space of an extended classical system. In some cases, this integral can be evaluated
by steepest descent,\cite{BenderBook} leading to a family of 
``instanton'' 
theories\cite{Perspective,Uses_of_Instantons,Milnikov2004,tunnel,
InstReview,Vaillant2018instanton,
Cvitas2018instanton,Erakovic2020excited,
asymtunnel}
that constitute rigorous asymptotic approximations to the exact quantum
result. Instantons correctly capture multidimensional
tunneling and zero-point energy effects but are 
relatively computationally inexpensive, since they only require local information
along the optimal tunneling pathway. 
This makes them an attractive
alternative to numerically exact 
techniques when studying complex molecular 
systems.\cite{water,hexamerprism,WaterChapter,formic,chiral,tropolone} 
Despite these advantages, instantons only yield approximate tunneling 
splitting values, whose accuracy deteriorates in low-barrier anharmonic 
systems.\cite{tunnel}

In this paper, we will focus on molecular systems with low barriers and anharmonic vibrations, such as methyl rotations, for which instanton theory is not appropriate.
Historically, studies of large-amplitude torsional
modes have helped make the connection between
electronic structure and conformational/reaction
energy barriers.
Spectroscopic features associated with methyl 
torsion remain a sensitive probe of intermolecular interactions in solids
and of rotation--vibration coupling in isolated molecules.
The low barriers can, however, pose a challenge. Methyl
groups connected to planar frameworks (aryl, nitryl, boryl)
exhibit barriers of only a few wavenumbers,\cite{Lister1978internal} 
for which instanton
approaches would fail drastically. The torsional barrier for the
methyl group in methanol is not quite as low (around \SI{350}{\per\cm}),
but still comparable to the harmonic torsional frequency of 
approximately \SI{290}{\per\cm}.\cite{Qu2013Methanol}
Such ``floppiness'' poses a problem not only for instantons but also
for wavefunction-based computations, since large-amplitude motions
must be represented with care to keep computational expense to a minimum.\cite{Carter2000multimodeRPH,Carrington2008ch5,Lauvergnat2014Methanol}

Developments in sparse-grid methods have pushed the limit
on the size of molecules for which full-dimensional wavefunction
calculations are feasible
(e.g., malonaldehyde in Ref.~\onlinecite{Lauvergnat2023malonaldehyde}), but further extensions to larger systems, especially ones with multiple large-amplitude modes, remain a challenge. Such systems can
be successfully tackled by standard instanton theory.\cite{hexamerprism,porphycene} However, instanton theory relies on a semiclassical approximation, and although its accuracy
can in some cases be improved with perturbative corrections,\cite{Lawrence2023PC} these are only applicable up to a point and will fail in extreme cases.
Nevertheless, strongly anharmonic systems with low barriers can be successfully simulated if one 
abandons the asymptotic approximation, choosing instead to evaluate
the (discretized) path integral nonperturbatively with numerical sampling methods. This brings us, then, to 
path-integral Monte Carlo (PIMC) and the main focus of our paper, path-integral
molecular dynamics (PIMD).\cite{Chandler+Wolynes1981,Parrinello1984PIMD,TuckermanBook,Ceperley1987exchange,
Ceperley1995PathIntegrals} 
Early work employed path-integral sampling to extract splitting 
values from low-temperature thermal density matrix elements
of model condensed-phase 
systems.\cite{Ceperley1987exchange,Ceperley1995PathIntegrals,
Alexandrou1988tunnelling,Marchi1991tunnelling} The approach was adapted 
by M\'{a}tyus and co-workers to study molecular rovibrational 
tunneling,\cite{Matyus2016tunnel1,Matyus2016tunnel2} and 
continued methodological developments have enabled its application
to water 
clusters.\cite{Vaillant2018dimer,Vaillant2019water,Zhu2022trimer}

To further expand the scope and accuracy of PIMD tunneling splitting
calculations, we will address two complications associated with
this approach.
First, the original PIMD procedure requires fitting simulation results
to a function not just of the tunneling splitting, but also
of a second ``tunneling time'' parameter.
At the very least, then, calculations must be conducted at two
different temperatures in order to obtain a value for the tunneling
splitting.
Second, M\'{a}tyus  et~al.\ accomplish rigorous projection 
onto states of definite angular 
momentum by performing numerical quadrature over rotations.
Hence, in principle, a complete calculation at one temperature
itself comprises multiple ``sub-calculations,''\footnote{By a single 
``sub-calculation'' we mean a complete thermodynamic integration
as described in \sref{ti}} one for every point on
the rotational quadrature grid. At the time of writing, we are not 
aware of
any applications to molecular systems that employ the full
procedure as outlined above.\footnote{A study by 
{\begin{NoHyper}\citeauthor{Vaillant2018instanton}
\end{NoHyper}}\cite{Vaillant2018instanton} 
did perform rotational projection by quadrature,
but based on instanton sub-calculations
rather than PIMD.}
Instead, previous PIMD calculations chose to reduce the computational 
cost by using only one orientation as defined by the instanton 
pathway.\cite{Matyus2016tunnel1,Vaillant2019water,Zhu2022trimer,
Vaillant2018dimer}
This neglects rovibrational coupling, which, although it may often be 
reasonably accurate, formally constitutes an uncontrolled 
approximation.

Here, we derive an alternative means of calculating tunneling
splittings that uses a modified PIMD formulation based on symmetrized partition
functions rather than density matrix elements.  The dynamics involve a new type of spring
that accounts for permutational and rotational symmetry, in an apparently
simple modification that eliminates both the ``tunneling time''
parameter and the numerical quadrature over orientations. This results in
a ``one-shot'' procedure that is formally exact and only calls for a single 
(thermodynamic 
integration) calculation to yield the tunneling splitting. We lay down 
the theoretical foundations
for the approach in \sref{theory}, which describes how to
extract tunneling splittings from symmetrized partition functions.
In \sref{practical} we give a brief summary of standard PIMD,
introduce the necessary modifications for dealing with symmetry projections and rotational
degrees of freedom, and summarize how these are used
to yield ratios of partition functions. \Sref{results} discusses
the numerical tests of the new rotational 
projection procedure on water and the application of our method
to splitting calculations in hydronium and methanol. \Sref{conclusions}
concludes the article.

\section{Theory%
\label{sec:theory}}
\subsection{Rotationless double-well systems%
\label{sec:rotationless}}

\begin{figure}[b]
\includegraphics{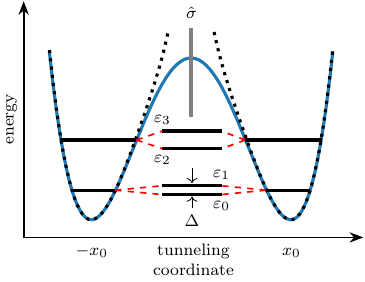}
\caption{A double-well potential (blue) with a mirror plane (segment 
drawn in gray) centered on the origin. The dashed lines denote the two
potential-energy surfaces in a hypothetical system with no tunneling
between the minima at $\pm x_0$. The corresponding unsplit
energy levels are shown as thick horizontal lines on the left and right 
sides of the plot. Thick horizontal lines in the center depict energy 
levels of the double well, for which tunneling splitting is observed. 
Small arrows indicate the ground-state splitting~$\Delta$.%
\label{fig:umbrella}
}
\end{figure}

Let us consider the tunneling splitting between the totally
symmetric ground state and a set of energy levels
that all have different symmetries and that are all below the first
excited totally symmetric state.\cite{Matyus2016tunnel2} For reasons explained in \sref{rotating} we refer
to systems with this property as ``rotationless.''
One such system is the symmetric one-dimensional 
double well, with quantum energy levels $\en_n$ (see~\fig{umbrella}).
In the low-temperature limit only the lowest-energy states contribute 
significantly to the partition function,
\begin{gather}\label{eq:ebasis_Z}
    Z = \sum_{n = 0}^{\infty} 
    \braket{n | \eu{-\beta \op{H}} | n} =
    \eu{-\beta \en_0} + \eu{-\beta \en_1} + \ldots \, .
\end{gather}
In order to extract the tunneling 
splitting, defined as $\Delta = \en_1 - \en_0$, we introduce 
the inversion operator $\op{\sigma}$, which acts on coordinates
as $\op{\sigma} x = -x$ and maps the two degenerate wells onto
each other. The ground state is symmetric under inversion,
while the first excited state is antisymmetric, meaning $\op{\sigma} \ket{0} = \ket{0}$ and $\op{\sigma} \ket{1} = -\ket{1}$.
We can use this property to obtain the low-temperature limit
of the ``symmetrized'' partition function,\cite{
FeynmanStatMech,Ceperley1995PathIntegrals}
\begin{gather}\label{eq:ebasis-Zsym}
    Z_{\sigma} = \sum_{n = 0}^{\infty} 
        \braket{n | \eu{-\beta \op{H}} \op{\sigma} | n} =
     \eu{-\beta \en_0} - \eu{-\beta \en_1} + \ldots \, .
\end{gather}
The tunneling splitting is then readily obtained from
\begin{gather}
\dfrac{Z-Z_{\sigma}}{Z+Z_{\sigma}} = 
\dfrac{1-\frac{Z_{\sigma}}{Z}}{1+\frac{Z_{\sigma}}{Z}} = \si{e}^{-\beta 
\Delta} \label{eq:dwell-Delta}
\end{gather}
in the limit as $\beta \to \infty$.

\subsection{Generalization to finite symmetries%
\label{sec:multi-well}}

The argument can be extended to multidimensional systems
with other symmetries in a similar manner to 
Ref.~\citenum{Matyus2016tunnel2}. Consider a Hamiltonian whose 
symmetries
form the group $\mathcal{G}$ of order $g$. Let $\op{P}$ be 
some symmetry operation in this group, for which we define
the symmetrized partition function 
\begin{equation}
    Z_P =  \Tr\mleft[ \eu{-\beta \Hop} \op{P} \mright].
\label{eq:defZ-P}
\end{equation}
As in \eq{eq:ebasis-Zsym}, we can expand the trace
in the energy eigenbasis,
\begin{align}
    Z_P & = \sum_{n} \sum_{l} \braket{nl | \eu{-\beta 
    \op{H}}\op{P} |nl},
\label{eq:ebasis-Za}
\end{align}
where $n$ ranges over all the energy levels, and $l$
ranges over the degenerate states comprising level $n$.
These states form an orthonormal basis for the irreducible 
representation (irrep)
$\Gamma^{(n)}$ and transform according to\cite{RHB}
\begin{align}
\op{P} \ket{nl} = \sum_{j} \ket{nj} D_{jl}^{(n)}(\op{P}).
\label{eq:P-transform}
\end{align}
Here $\mat{D}^{(n)}(\op{P})$ is a representation matrix, whose trace
\begin{align}
\chi^{(n)}_{P} = \sum_{j} D_{jj}^{(n)}(\op{P}) \label{eq:character}
\end{align}
is known as the character of $\op{P}$ in $\Gamma^{(n)}$.
By virtue of \eqs{eq:P-transform}{eq:character}, and the orthonormality 
of $\ket{nl} $, we can rewrite \eqn{eq:ebasis-Za} as
\begin{align}
   Z_P & = \sum_{n,l,j} 
 \braket{nl | \eu{-\beta \op{H}} |nj}  D_{jl}^{(n)}(\op{P}) \nonumber \\
 & = \sum_{n,l,j}  \eu{-\beta \en_n} \delta_{lj}  D_{jl}^{(n)}(\op{P}) = \sum_{n} \eu{-\beta \en_n} \chi^{(n)}_{P}.
 \label{eq:Za-char}
\end{align}
Note that if two or more operations belong to the same conjugacy class 
$\alpha$ of order $g_{\alpha}$, they have identical
characters.\cite{BunkerBook,RHB} It is, therefore, sufficient to 
calculate 
$Z_P = Z_{\alpha}$ 
for a single member of a conjugacy class. 
To extract 
individual energy levels from \eqn{eq:Za-char} we can, in general, use
the character orthogonality relation
\begin{equation}
\sum_{\alpha} g_{\alpha} \chi_{\alpha}^{(m)*} \chi_{\alpha}^{(n)}  = g 
\, \delta_{\Gamma^{(m)} \Gamma^{(n)}},
\label{eq:chi-ortho}
\end{equation}
where summation is over all the conjugacy classes of~$\mathcal{G}$, 
$\chi_{\alpha}^{(m)*}$ denotes the complex conjugate of 
$\chi_{\alpha}^{(m)}$, and $ \delta_{\Gamma^{(m)} \Gamma^{(n)}} $ is
$1$ if energy levels $m$ and $n$ correspond to the same irrep,
and $0$ otherwise. From \eqn{eq:chi-ortho} it follows that
\begin{equation}\label{eq:chi-ortho-2}
\sum_{\alpha} \frac{g_{\alpha}}{g} \chi_{\alpha}^{(m)*} 
Z_\alpha
= \sum_{n} \eu{-\beta \en_n} \delta_{\Gamma^{(m)} \Gamma^{(n)}}
= \eu{-\beta \en_m}
\end{equation}
for $\beta \to \infty$, with $\en_m$ denoting the lowest energy
level whose states transform as $\Gamma^{(m)}$. The tunneling
splitting pattern ($\Delta_{m,n} = \en_m - \en_n$) may, therefore, be 
deduced by evaluating expressions of the form
\begin{equation}
\eu{-\beta \Delta_{m,n}} = \frac{
 \sum_{\alpha} g_{\alpha} \chi_{\alpha}^{(m)*} Z_{\alpha}
}{
 \sum_{\alpha} g_{\alpha} \chi_{\alpha}^{(n)*} Z_{\alpha}
}, \label{eq:Delta-mn}
\end{equation}
using tabulated values of $g_{\alpha}$ and $\chi_{\alpha}^{(m)}$.
The expression includes $Z_{\alpha}$ for every conjugacy class,
but typically one need only
calculate a subset of those, as illustrated in the following example.

 \begin{table}[t]
 \caption{
 Character table for the $C_{3v}$ point group.
 Column headings refer to conjugacy classes, with leading integers giving the 
 corresponding order $g_{\alpha}$, understood to be $1$ if not stated explicitly.
 Conventional irreducible representation labels are used as row headings. 
 \label{tab:c3}
 }
 \renewcommand{\arraystretch}{1.5}
 \begin{ruledtabular}
 \begin{tabular}{>{\hspace*{2ex}}l|*{3}{r}<{\hspace*{1em}}}
 $C_{3v}$ & $E$  & $2C_3$\hspace*{-1ex} & $3\sigma_v$\hspace*{-1ex} \tabularnewline
  \midrule
  $A_1$ & $1$ & $ 1$ & $ 1$ \tabularnewline
  $A_2$ & $1$ & $ 1$ & $-1$ \tabularnewline
  $E$   & $2$ & $-1$ & $ 0$ \tabularnewline
 \end{tabular}
 \end{ruledtabular}
 \end{table}

\begin{figure}[b]
\includegraphics{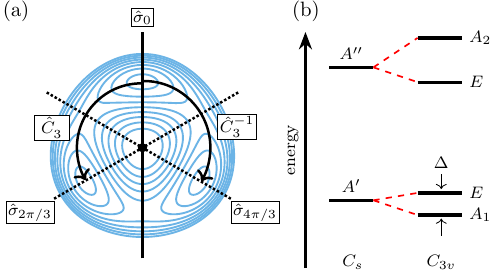}
\caption{(a)~Contour plot of a model potential with $C_{3v}$ symmetry.
The solid black line denotes the symmetry plane of the reference 
non-tunneling system
confined to the top well, described by the $C_s$ point group. The full 
$C_{3v}$ group includes two further symmetry planes (dotted black 
lines) and two
three-fold rotations (curved arrows). (b)~The tunneling splitting 
pattern for the $A'$ ground state and the $A''$ vibrationally excited
state of the reference system (the energy diagram does not include any 
excitations in the radial mode). Thick black lines denote the energy 
levels before
($C_s$) and after ($C_{3v}$) tunneling splitting.%
\label{fig:c3v-pot}}
\end{figure}

Consider the $C_{3v}$ point group, which describes the symmetry of a 
$(2\pi/3)$-periodic potential, sketched in \fig{c3v-pot},
and whose characters are given in \tref{c3}. 
$C_{s}$~is a subgroup of $C_{3v}$ describing the feasible
symmetries\cite{LonguetHiggins1963symmetry} of the hypothetical 
non-tunneling version of the system confined to a
single potential-energy well (our ``reference system'').
Following {\begin{NoHyper}\citeauthor{BunkerBook}\end{NoHyper}},\cite{BunkerBook} 
we determine the representation in $C_{3v}$ that is induced by the 
irrep of a reference state in $C_{s}$.
For the ground state we must consider the 
totally symmetric irrep ($A'$ in the $C_{s}$ point group). Its 
induced representation in $C_{3v}$ is $A_{1} \oplus E$,
which gives the symmetries of the tunneling-split states, namely, 
a totally symmetric $A_1$ ground state and a doubly degenerate $E$ excited state.
There is, therefore, just one tunneling splitting to consider,
$\Delta_{1,0} \equiv \Delta$, given by \eqn{eq:Delta-mn} as
\begin{align}
	\label{eq:C3-Delta-0}
	\eu{-\beta \Delta} & = \frac{
		2 Z - 2 Z_{C_3}
	}{
		Z + 2 Z_{C_3} + 3 Z_{\sigma_v} }.
\end{align}
Taking the low-temperature limit of \eqn{eq:Za-char}, we find that
\begin{subequations}
	\label{eq:c3v-partfs}
	\begin{align}
		\label{eq:c3v-partf-E}
		Z & = \eu{-\beta \en_{\!A_1}} + 2 \mkern1mu \eu{-\beta 
		\en_{E}}, \\
		\label{eq:c3v-partf-C3}
		Z_{C_3} & = \eu{-\beta \en_{\!A_1}} - \eu{-\beta \en_{E}}, \\
		\label{eq:c3v-partf-sigma}
		Z_{\sigma_v} & = \eu{-\beta \en_{\!A_1}},
	\end{align}
\end{subequations}
where $\en_{\Gamma}$ denotes the lowest energy level that transforms according to
the irreducible representation $\Gamma$. \Eqn{eq:c3v-partfs} does not
contain any terms in $\en_{\!A_2}$, since, based on our symmetry analysis, 
this energy derives from an excited vibrational state of the reference 
system. This
is assumed to be well separated from the tunneling-split ground state, 
such that the Boltzmann factor $\eu{-\beta \en_{\!A_2}}$ becomes
negligible compared to the other terms in \eqn{eq:c3v-partfs} as $\beta \to \infty$.

Given this, there are several possible ways to derive a
simplified expression for $\Delta$. The first possibility
is to re-evaluate \eqn{eq:c3v-partfs} in the smallest subgroup
that still assigns different symmetry labels to
the different energy levels of the tunneling-split
ground state. Here, the appropriate subgroup is $C_3$.
The second possibility is to express one of the symmetrized partition 
functions in terms of the others, 
e.g., $Z_{\sigma_v} = \tfrac{1}{3}[Z + 2 Z_{C_3}]$, followed by 
substitution
into \eqn{eq:C3-Delta-0}. Yet another option is to pick
as many expressions as there are non-negligible Boltzmann factors,
e.g., \eqs{eq:c3v-partf-E}{eq:c3v-partf-C3}, 
and to solve the simultaneous linear equations
for these factors (and, hence, for the tunneling splittings) 
directly. 
The three approaches all give the same result,\footnote{Alternative 
expressions in term of $Z$ and $Z_{\sigma_v}$ or $Z_{\sigma_v}$ and 
$Z_{C_3}$ can be derived, but ours is the better choice for numerical 
calculations.} which for our example reads
\begin{align}
\eu{-\beta \Delta} & =
 \frac{
  Z - Z_{C_3}
 }{
   Z + 2 Z_{C_3} } =  \frac{
     1 - \frac{Z_{C_3}}{Z}
    }{
      1 + 2 \frac{Z_{C_3}}{Z} },
 \label{eq:C3-Delta}
\end{align}
and is subtly different from the simple double-well case described
by \eq{eq:dwell-Delta}. The general procedure for obtaining such 
expressions can be summarized
as follows:
\begin{enumerate}
	\item Determine the symmetries of the tunneling-split states.
	\item Obtain the low-temperature limit of \eqn{eq:Za-char}
	for a subset of all conjugacy classes, keeping only the Boltzmann
	factors that correspond to the states in step 1.
	\item Once the set contains as many linearly independent expressions
	as the number of levels in step 1, solve for the Boltzmann factors.
	\item Express the tunneling splitting constant(s) in terms of the
	symmetrized partition functions by taking ratios of the expressions
	in step 3.
\end{enumerate}
For this introductory example we have considered a point group, 
such as one might encounter in discussions of rigid molecules.
However, the results of this section also apply to
non-rigid molecules, for which the $\op{P}$ operators represent 
the permutations of indistinguishable atoms and inversion through the
molecular center of mass (CoM). Feasible combinations of these 
operations
belong to the molecular symmetry group (MS),\cite{
LonguetHiggins1963symmetry} which replaces the point group
in subsequent sections. For completeness, we note that
it is with the help of the MS group that nuclear
spin-statistical weights can be assigned 
to tunneling-split energy levels.\cite{BunkerBook}

\subsection{Rotating systems%
\label{sec:rotating}}

In addition to discrete symmetries of the kind described in \sref{rotationless}, a
molecular Hamiltonian is invariant under rotation.\footnote{It is also 
invariant under translation, but the translational contribution is 
exactly factorizable and can, therefore, be easily 
removed\cite{BunkerBook}} Consequently, its eigenfunctions can 
be labeled by two further quantum numbers: the total angular momentum 
quantum number, $J$, and the projection quantum number, $M$. The latter 
gives the projection of the angular momentum onto the space-fixed 
$z$-axis and ranges in integer steps from $-J$ to $J$. 
As we consider systems in the absence 
of external electromagnetic fields,
states that only differ by the value of $M$ are 
degenerate. 

The energy separation between states of different
$J$ can in some cases be similar to or even smaller
than a rovibrational
tunneling splitting. This means that simply taking the low-temperature
limit of \eqn{eq:Delta-mn}, as outlined in 
\srefs{rotationless}{multi-well}, will not be sufficient
to isolate contributions from the tunneling-split ground state.
We must instead tackle this problem by explicitly
selecting for a particular $J$, 
i.e., projecting onto states with a specific total angular momentum.

From here on we assume that the molecule in question is closed-shell, so there are
no interactions between rotational angular momentum and electron spin. We also neglect 
any hyperfine structure due to coupling to nuclear spin. Under these assumptions, we can 
denote the eigenfunctions of the molecular Hamiltonian as $\ket{n,J,M}$, where $n$ is shorthand for any other relevant quantum numbers or symmetry labels. These eigenfunctions
transform as
\begin{equation}
	\op{R}(\bm{\Omega}) \ket{n, J, M} = \sum_{L=-J}^{J} \ket{n, J, L} 
	D_{L M}^{(J)}(\bm{\Omega}),
\end{equation}
where $\op{R}(\bm{\Omega})$ is a rotation specified by $\bm{\Omega}$
(which could be a set of Euler angles, 
or a unit quaternion, or any other valid 
representation), and $D_{LM}^{(J)}(\bm{\Omega})$ is a Wigner 
$D$-matrix.\cite{zare1988angular} The latter satisfy an orthogonality 
relation
\begin{equation}
	\label{eq:D-ortho}
	\int \! \rmd \bm{\Omega} \, D_{L'  M'}^{(J')}(\bm{\Omega})^{*}
	D_{L M}^{(J)}(\bm{\Omega}) = \delta_{J' J} \delta_{L' L}
	\delta_{M' M} \frac{8 \pi^2}{2J+1},
\end{equation}
which can be used to construct a projection onto a state with a specified $J$.
By analogy with \sref{rotationless}, we define a ``rotated'' partition function
\begin{subequations}
\begin{align}
	Z(\bm{\Omega}) & = \Tr \mleft[ \eu{-\beta \op{H}} \op{R}(\bm{\Omega}) \mright] \\ 
	               & = \sum_{n,J,M} \braket{
		n, J, M | \eu{-\beta \hat{H}} \hat{R}(\bm{\Omega}) | n, J, M} \\
 	               & = \sum_{\substack{n,J\\M,L}} 
 	               D_{LM}^{(J)}(\bm{\Omega})
	\braket{
		n, J, M | \eu{-\beta \hat{H}} | n, J, L}
\end{align}
\end{subequations}
and use \eqn{eq:D-ortho} to get
\begin{subequations}
\label{eq:rot-proj}
\begin{align}
	& \int  D_{M'M'}^{(J')}(\bm{\Omega})^{*} Z(\bm{\Omega}) \, \rmd 
	\bm{\Omega} \nonumber \\
	& \qquad {} = \frac{8 \pi^2}{2J' + 1} \sum_{n} 
	\braket{
		n, J', M' | \eu{-\beta \hat{H}} | n, J', M'}
	\label{eq:rot-proj-sum} \\
	& \qquad {} = \frac{8 \pi^2}{(2J' + 1)^2} \, Z^{J = J'}. \label{eq:rot-proj-def}
\end{align}
\end{subequations}
On going from \eqn{eq:rot-proj-sum} to \eqn{eq:rot-proj-def} we recall the 
$(2J' + 1)$-fold degeneracy of the states that only differ by the 
projection quantum number
and recognize that the resulting sum is the contribution to the total 
partition function from states with angular momentum $J'$.
Given that \mbox{$D^{(0)}_{00}(\bm{\Omega}) = 1$}, \eqn{eq:rot-proj} can be rearranged to read
\begin{equation}
	\label{eq:rot-partf-J}
	Z^{J=0} = \frac{1}{8 \pi^2} \int  \Tr[\eu{-\beta \op{H}} \op{R}(\bm{\Omega})] \, \rmd \bm{\Omega},
\end{equation}
i.e., states with $J=0$ can be isolated by taking an unweighted average over all ``rotated'' partition functions.

Having treated rotational symmetries, we can extract ground-state 
tunneling splitting constants using the same approach as in 
\sref{multi-well}. In a straightforward generalization, we define 
\begin{subequations}
	\label{eq:rot-partf-P}
\begin{align}
	\label{eq:rot-partf-aW}
	Z_{P}(\bm{\Omega}) & =  \Tr\mleft[ \eu{-\beta \Hop} \op{P} \op{R}(\bm{\Omega}) \mright], \\
	\label{eq:rot-partf-aJ}
	Z^{J=0}_{P} & = \frac{1}{8 \pi^2} \int Z_{P}(\bm{\Omega})  \, \rmd \bm{\Omega}.
\end{align}
\end{subequations} 
Given a $ Z_{\alpha}^{J = 0} $ for every conjugacy class, $\alpha$, of 
the MS group (or a subgroup thereof), \eqn{eq:Delta-mn} directly yields
the desired splitting constants.\footnote{
As an aside, we note that transitions between rovibronic states
with $J = 0$ are forbidden.\cite{BunkerBook} However, the corresponding 
tunneling splitting can be extracted from experimental data by fitting to a spectroscopic Hamiltonian, as in Ref.~\citenum{Xu1997MeOHsplit}
}

\section{Methods%
\label{sec:practical}}

\subsection{Path-Integral Molecular Dynamics%
\label{sec:pimd}}

For \eqn{eq:Delta-mn} to be of practical use, we need a means of 
calculating partition functions without having to solve the 
Schr\"{o}dinger equation. The path-integral
formulation of quantum mechanics is particularly suited to this purpose 
and is easily extended to obtain the symmetrized
partition function, $Z_P$. 
First, we apply the symmetric Trotter product 
formula,\cite{Suzuki1985Split}
\begin{align}
	\eu{-\beta \op{H}} = \lim_{N\rightarrow\infty} \mleft( 
	\eu{-\beta_N\op{V}/2} \mkern1.5mu 
	\eu{-\beta_N \op{T}}  \mkern1.5mu
	\eu{-\beta_N\op{V}/2} \mright)^N ,
\end{align}
where $\beta_N = \beta/N$, $\op{V}$ is the potential energy operator and $\op{T}$ is the kinetic energy operator. Denoting the bracketed product of operators as
$\op{O}_N$, in the limit as $N \to \infty$ we have
\begin{align}
	Z_{P} &  = \int \! \rmd \vec{x}_1 \,
	\braket{\vec{x}_1 |  ( \op{O}_N )^N \op{P} | \vec{x}_1} \nonumber \\
	&  = \int \! \rmd \cvec{x} \, \braket{\vec{x}_{N} |  \op{O}_N 
	\op{P} | \vec{x}_{1}} \times
	 \prod_{i = 1}^{N}  \braket{\vec{x}_{i} |  \op{O}_N  | 
	 \vec{x}_{i+1}}.
	 \label{eq:Z-trot}
\end{align}
To arrive at the right-hand side of \eqn{eq:Z-trot},
we inserted $N-1$ position resolutions of the identity. The bold 
cursive $\cvec{x}$
refers to the set of $N$ vectors $\{ \vec{x}_i \, | \, 
i = 1 \ldots N \}$, where $\vec{x}_i  = ( x^{\smash{(i)}}_1, \, 
x^{\smash{(i)}}_2, \, 
\ldots, \, x^{\smash{(i)}}_f )$ and $f$ is the number of configurational
degrees of freedom (DoF). For the 
integration, we use the shorthand notation
\begin{equation}
	\int \rmd \cvec{x} \equiv \prod_{i = 1}^N \mleft[ \int \rmd \vec{x}_i \mright].
\end{equation}
Upon evaluating
the matrix elements in \eqn{eq:Z-trot}, we can express the partition function as
\begin{subequations}
	\label{eq:rp-main}
	\begin{align}
		\label{eq:rp-partf}
		Z_P & = (2 \pi \hbar)^{-N f} \int \rmd \cvec{p}
		\int \rmd \cvec{x} \ \eu{-\beta_N \funOpN[P]{H}(\cvec{p}, 
		\cvec{x})} \\
		\label{eq:prp-ham}
		\funOpN[P]{H} & = \funOpN[(\text{open})]{H} + \sum_{k=1}^{f} 
			\frac{m_k \omega_N^2}{2} \mleft[x^{(N)}_{k} \! - \op{P} 
			x^{(1)}_{k}\mright]^2  \\
		\label{eq:orp-ham} &
		\hspace*{-1.5em}
		\begin{multlined}[t][0.775\columnwidth]
		\funOpN[(\text{open})]{H}  = \sum_{i = 1}^N \mleft[
					\sum_{{k} = 1}^{f} \frac{(p^{(i)}_{k})^2}{2 m_{k}} 
					+ 
					V(\vec{x}_i) \mright] \\
		 {} + \sum_{i=1}^{N-1} \sum_{k=1}^f \frac{m_k 
			\omega_N^2}{2} 
			\mleft[x^{(i+1)}_{k} \! - 
			x^{(i)}_{k}\mright]^{2},
		\end{multlined}
	\end{align}
\end{subequations}
where $m_{k}$ is the mass of the $k$-th DoF
and $\cvec{p}$ refers to the set of momenta $\{ \vec{p}_i \, | \, 
i = 1 \ldots N \}$.
The Hamiltonian in this expression describes
an extended \emph{classical} system
comprised of $N$ replicas of the original
set of particles that reside on the potential 
energy surface $V(\vec{x})$. Replicas with adjacent indices
($i$ and $i \pm 1$) are connected by harmonic springs
of frequency $\omega_N = 1 / \beta_N \hbar$,
so that the term $\funOpN[\smash{(\text{open})}]{H}$ 
corresponds to a linear polymer or an open necklace
in which replicas are the ``beads.''

The second term in \eqn{eq:prp-ham} introduces a spring between
the ends of the necklace (beads 1 and $N$). When the symmetry
operator $\op{P}$ is taken to be the identity $\op{E}$, 
$\op{E} \vec{x} = \vec{x}$, the additional spring connects
the extended classical system into a ``ring polymer.''
In this case, $Z_E$ refers to the standard quantum partition function, 
$Z$, and \eqn{eq:rp-partf} represents the discretized version of a 
Feynman path 
integral expression for the trace of the Boltzmann 
operator.\cite{Feynman} Crucially, \eqn{eq:rp-partf}
is in the form of a classical partition function 
(for the ring polymer at a temperature $T \times N$). As such, it can be evaluated using 
well-established techniques from classical statistical mechanics.
Because of the connection to the path-integral 
formalism, simulations that sample \eqn{eq:rp-partf}
using classical dynamics generated by $\funOpN[\smash{P}]{H}$
(in combination with a suitable thermostat to ensure 
canonical sampling) are known as path-integral molecular 
dynamics 
(PIMD).\cite{TuckermanBook} Many PIMD techniques are easily
compatible with other choices of $\op{P}$, for which the cyclic 
symmetry of the ring polymer 
is broken.
As we shall see in \srefs{h3o}{h3coh},
the ends of the necklace, $\vec{x}_1$ and 
$\vec{x}_N$, tend to explore
degenerate minima related by the symmetry operation $\op{P}$, and the 
intermediate
beads sample the potential energy barrier connecting the two low-energy 
regions. 

We note in passing that the symmetrized partition functions used here
also arise in the context of bosonic and fermionic 
quantum statistics.\cite{FeynmanStatMech}
For example, \eqn{eq:chi-ortho-2} evaluated for the
totally symmetric irrep of the complete nuclear
permutation-inversion (CNPI) group\cite{BunkerBook}
is the contribution to the
bosonic partition function from the states that have a totally symmetric nuclear spin part;
evaluated for the antisymmetric irrep, it gives the contribution from the states with an antisymmetric nuclear spin part.
The contributions to the fermionic partition function can be obtained in an equivalent way.\footnote{Here, we are
not required to subtract terms of a similar 
magnitude, and so do not suffer from the fermionic sign problem.}
Other irreps could be 
paired with nuclear spin states of different symmetries
or ascribed to hypothetical particles
with more exotic quantum statistics.
Due to this formal connection, we could in principle
make use of PIMC\cite{Ceperley1995PathIntegrals} and PIMD methods 
\cite{Runeson2018Fermions,Hirshberg2019bosons} developed for simulating 
the quantum thermodynamics of bosonic and fermionic systems.
For now, however, it seems more pragmatic to use the PIMD
formulation presented in this section, combined with 
thermodynamic integration (TI) as described in \sref{ti}.

\subsection{Eckart Spring%
\label{sec:eck-spring}}

The distinguishing feature of the $J=0$ partition function in
\eqn{eq:rot-partf-J} is the additional integral over orientations, $\bm{\Omega}$. 
Repeating the manipulations from \sref{pimd} results in the following
``polymer'' partition function:
\begin{align}
	& Z^{J=0} = (2 \pi \hbar)^{-N f} \frac{1}{8 \pi^2}
	\int \rmd \cvec{p} \int \rmd \cvec{r} \, \mleft\{ 
	\vphantom{\sum_M^N}\eu{-\beta_N \funOpN{H} (\cvec{p}, \cvec{r})} 
	\times  \mright. \nonumber \\
	& \!\! \mleft.
	\int \! \rmd \bm{\Omega} \, \exp \mleft[-\frac{\beta_N \omega_N^2}{2} 
	\sum_{\mathclap{a=1}}^{\nat} m_{a} \mleft \lVert 
		\vec{r}^{(N)}_{a} \! - \op{R}(\bm{\Omega})\vec{r}^{(1)}_{a} 
	\mright\rVert^2 \mright] \! \mright\},
	\label{eq:rp-rot-partf}
\end{align}
where $\nat$ is the number of atoms in the molecule, \mbox{$ f = 3 
\nat$,} and 
$\vec{r}_{a}^{\smash{(i)}}$ is a three-dimensional coordinate vector $\big(
	x_{a}^{\smash{(i)}}\!, \, y_{a}^{\smash{(i)}}\!, \,	z_{a}^{\smash{(i)}} \big)$
describing the $a$-th atom of mass $m_{a}$ in the $i$-th
system replica (bead).
The integral over $\bm{\Omega}$ can be expressed
as
\begin{equation}\label{eq:Omega-int}
I(\bm{r}) = \frac{1}{8 \pi^2} \int \rmd \bm{\Omega} \, 
\exp\left[-\frac{N 
f\big(\bm{\Omega}; \vec{r}^{(N)}\!, \vec{r}^{(1)}\big)}{2\beta \hbar^2}\right],
\end{equation}
where
\begin{equation}\label{eq:norm2}
f\big(\bm{\Omega}; \vec{r}^{(N)}\!, \vec{r}^{(1)}\big) = 
\sum_{\mathclap{a=1}}^{\nat} m_{a}
	\mleft \lVert 
	\vec{r}^{(N)}_{a} \! - 
	\op{R}(\bm{\Omega})\vec{r}^{(1)}_{a} 
	\mright\rVert^2.
\end{equation}
For the purposes of evaluating \eqn{eq:Omega-int},
we can approximate $f\big(\bm{\Omega}; \vec{r}^{(N)}\!, 
\vec{r}^{(1)}\big)$ by a Taylor series expansion in $\bm{\Omega}$
about its minimum, truncated at second order. In doing so, we replace 
the (rather non-trivial) 
integral over $\bm{\Omega}$ with a three-dimensional Gaussian integral, which can be
evaluated analytically. It can be shown that the relative error 
introduced by this approximation vanishes in the many-bead ($N \to 
\infty$) limit, which is the same limit
as we generally aim for in PIMD simulations. This way of approximating 
integrals is known as Laplace's method
or, more generally, the method of steepest descent.\cite{BenderBook} It 
is a standard technique of 
asymptotic analysis and forms the foundation of instanton 
theory.\cite{InstReview}
However, unlike in instanton theory, the approximation here can be made 
exact by increasing $N$. The error
scales as $O(N^{-1})$, in principle decaying more slowly
than the $O(N^{-2})$ error from ring-polymer 
discretization.\cite{Schmidt1995Trotter}
Despite the formally slower scaling, we observe that the
Eckart spring does not negatively impact the rate at which
tunneling splitting calculations converge with respect
to the number of beads.

For the steepest-descent approach to be practical, we need an efficient 
way of locating the minimum of $f(\bm{\Omega})$ (from here on, the
parametric dependence on $\vec{r}^{(N)}\!$ and $\vec{r}^{(1)}\!$ is
dropped for brevity). The problem amounts to rotating replica 1 about 
its CoM such that
the mass-weighted square distance to replica $N$ is minimized. This 
concept is
closely tied to the Eckart frame, which is central to 
the construction of the rovibrational Hamiltonian\cite{Eckart1935,BrightWilsonBook} and has previously
featured in another PIMD method developed by one of the 
authors.\cite{Trenins2019} Note, however, that the Eckart
frame is not used to define a set of curvilinear coordinates
in the present work; all our expressions are in a Cartesian frame.

The algorithm for finding
the Eckart rotation is comprehensively laid out 
in Ref.~\onlinecite{Krasnoshchekov2014Eckart}. Here, we summarize the 
key details of the procedure. 
First, the rotation is represented by a unit quaternion $\vec{q} = (q_0, \, q_1, \, q_2, \, q_3)$
with a real part $q_0$. This is related to the rotation matrix via
\begin{equation}
	\label{eq:qua2mat}
	\begingroup
	\renewcommand*{\arraystretch}{1.5}
	\mat{R}(\bm{\Omega}) = 
	\begin{pmatrix}
	1 - 2 q_2^2 - 2 q_3^2 & 2 q_1 q_2 + 2 q_0 q_3 & 2 q_1 q_3 - 2 q_0 q_2 \\
	2 q_1 q_2 - 2 q_0 q_3 & 1 - 2 q_1^2 - 2 q_3^2 & 2 q_2 q_3 + 2 q_0 q_1 \\
	2 q_1 q_3 + 2 q_0 q_2 & 2 q_2 q_3 - 2 q_0 q_1 & 1 - 2 q_1^2 - 2 q_2^2 \\
	\end{pmatrix},
	\endgroup
\end{equation}
such that
\begin{equation}
	\label{eq:rot-def}
	\op{R}(\bm{\Omega}) \vec{r}_a^{(1)} = \mat{R}(\bm{\Omega}) \mleft[
		\vec{r}_a^{(1)} - \vec{r}^{(1)}_{\mathrm{com}}
	\mright] + \vec{r}^{(1)}_{\mathrm{com}}
\end{equation}
with
\begin{equation}
	\label{eq:com-def}
	\vec{r}^{(1)}_{\mathrm{com}} = \sum_{a = 1}^{\nat} m_{a} 
	\vec{r}_{a}^{(1)} 
	\Big / \sum_{a = 1}^{\nat} m_{a}
\end{equation}
denoting the CoM of replica 1. For the rest of this section it is assumed that
we are working in the CoM frame of replica 1, such that 
$\vec{r}^{\smash{(1)}}_{\mathrm{com}} = 
\vec{0}$.
 With the definitions in \eqt{eq:qua2mat}{eq:com-def},
it can be shown that the function in \eqn{eq:norm2} is a quadratic form 
in $\vec{q}$,
\begin{equation}
	f(\bm{\Omega}) = \vec{q} \cdot \mat{C} \cdot \vec{q},
	\label{eq:quad-form}
\end{equation}
with $\mat{C}$ a $4 \times 4$ real symmetric matrix that is a simple 
function of $m_{a}$, $\vec{r}_{a}^{\smash{(1)}}$ and $\vec{r}_{a}^{\smash{(N)}}$ 
($a = 1,\dotsc,\nat$). The explicit form of $\mat{C}$ is given in 
Eq.~(24) of Ref.~\onlinecite{Krasnoshchekov2014Eckart} where, in our 
case,
\begin{align}
x_{+a} & = x_{a}^{(N)} + x_{a}^{(1)}, & 
x_{-a} & = x_{a}^{(N)} - x_{a}^{(1)},
\end{align}
and similarly for $y$ and $z$ coordinates. The 
quadratic form can be diagonalized by finding the
eigenvalues and eigenvectors of $\mat{C}$. The smallest eigenvalue 
$\lambda_{\smash{0}}$ 
corresponds to the minimum of $f(\bm{\Omega})$, and the associated 
normalized
eigenvector $\vec{q}^{\smash{(0)}}$ is the quaternion that accomplishes 
the rotation onto the minimum. In our case, the
ring-polymer potential ensures that $\vec{r}^{(N)}\!$ and $\vec{r}^{(1)}\!$ have a strong structural similarity and can be 
rotated onto each other such that they are nearly coincident. The similarity of the structures means
that the optimal rotation is unique, unless
the molecules fluctuate about a (nearly) linear 
geometry.\cite{Jorgensen1978EckFrame}
 Construction and diagonalization of 
$\mat{C}$ are numerically inexpensive relative to the calculation
of the PES, so the outlined procedure can be readily incorporated
into a PIMD simulation.

To proceed with the steepest-descent approximation, we denote the optimal rotation 
with $\mat{R}(\eckRot)$, and the corresponding rotated structure
with $ \eckR{}^{\smash{(1)}}_{a} = \mat{R}(\eckRot) 
\vec{r}^{\smash{(1)}}_{a} $.
Using $\mat{R}(\bm{\Omega}) = \mat{R}(\delta \bm{\Omega}) 
\mat{R}(\eckRot)$,
the function $ f(\bm{\Omega})$ is rewritten as
\begin{equation}
    \begin{multlined}[t][0.75\columnwidth]
	f(\bm{\Omega}) = \sum_{\mathclap{a=1}}^{\nat} m_{a} \mleft \lVert 
	\vec{r}^{(N)}_{a} \! - \eckR^{(1)}_{a} 
	\mright\rVert^2 \\
	 {} + 2 \sum_{a=1}^{\nat} m_{a}
	 \vec{r}_{a}^{(N)} \cdot  [\mat{I} - \mat{R}(\delta \bm{\Omega})] 
	 \cdot 
	 \eckR_{a}^{(1)},
	 \end{multlined}
\end{equation}
where $\mat{I}$ is the $3 \times 3$ identity matrix. Next we express 
the rotation 
$\mat{R}(\delta \bm{\Omega})$
about the minimum in terms of Tait--Bryan angles 
$\begin{pmatrix} \phi & \theta & \chi \end{pmatrix}$,
\cite{Goldstein} 
\begin{align}
	\mat{R}(\delta \bm{\Omega}) & = 
	\begin{pmatrix}
		c_{\theta} c_{\chi} & -c_{\theta} s_{\chi} & s_{\theta} \\
		s_{\phi} s_{\theta} c_{\chi} + c_{\phi} s_{\chi}  &
		-s_{\phi} s_{\theta} s_{\chi} + c_{\phi} c_{\chi}  &
		-s_{\phi} c_{\theta} \\
		-c_{\phi} s_{\theta} c_{\chi} + s_{\phi} s_{\chi}  &
		c_{\phi} s_{\theta} s_{\chi} + s_{\phi} c_{\chi}  &
		c_{\phi} c_{\theta}
	\end{pmatrix} ,
\end{align}
where $c_\phi = \cos \phi$, $s_\phi = \sin \phi$, etc. The 
corresponding integral over 
orientations is
\begin{align}
	\int \rmd \bm{\Omega} & = \int_{-\pi}^{\pi} \rmd \phi \int_{-\pi /2}^{\pi /2} \cos \theta \, \rmd \theta \int_{-\pi}^{\pi} \rmd \chi \nonumber \\
	& = \int_{-\pi}^{\pi} \rmd \phi \int_{-1}^{1} \rmd z \int_{-\pi}^{\pi} \rmd \chi,
\end{align}
with $z = \cos \theta$. Now we expand $\mat{I} - \mat{R}(\delta 
\bm{\Omega})$ in a Taylor
series about $\left( \phi,\, z,\, \chi \right) = \vec{0}$. 
Up to second order,
\begin{equation} \label{eq:rotmat-taylor}
    \begin{multlined}[t][0.85\columnwidth]
    \mat{I} - \mat{R}(\delta \bm{\Omega}) \sim
		\begin{pmatrix}
			0 & \chi & -z \\
			- \chi & 0 & \phi \\
			z  & -\phi  & 0
		\end{pmatrix} \\
    \rule{0pt}{2.5em}
	  + \frac{1}{2}\begin{pmatrix}
			z^2 + \chi^2 & 0 & 0 \\
			-2 z \phi  & \phi^2 + \chi^2 & 0 \\
			- 2 \phi \chi &  - 2 z \chi & z^2 + \phi^2
		\end{pmatrix}.		
	\end{multlined}
\end{equation}
The first (linear) term of \eqn{eq:rotmat-taylor} contributes
\begin{equation}
	2 \sum_{a=1}^{\nat} m_{a}  \vec{r}_{a}^{(N)} \cdot
	\begin{pmatrix}
		0 & \chi & -z \\
		- \chi & 0 & \phi \\
		z  & -\phi  & 0
	\end{pmatrix} \cdot \mkern2mu \eckR_{a}^{(1)}
\end{equation}
to $f(\bm{\Omega})$. This evaluates to zero, because $\eckR_a^{(1)}$
satisfies the rotational Eckart conditions.\cite{BrightWilsonBook} 
Hence, 
only the second- (and higher-) order terms make a non-zero contribution 
around the stationary point
\begin{equation}
	f(\bm{\Omega}) \sim \sum_{a=1}^{\nat} m_{a} 
	\big \lVert \mathbf{r}_{a}^{(N)}
	- \eckR_{a}^{(1)} \big \rVert^2 + \delta \bm{\Omega}
	\cdot \mat{\eckmom}
	\cdot \delta \bm{\Omega},
\end{equation}
where $\delta \bm{\Omega}$ stands for 
$\left(\phi,\, z,\, \chi\right)$ and $\mat{\eckmom} = 
\mat{\eckmom}[\vec{r}^{(N)}\! , 
\vec{r}^{(1)} ]$ is a $3 \times 3$ 
matrix with elements
\begin{align}
	\label{eq:mat-A}
	\eckmom_{ij}[\vec{r}^{(N)}\! , 
\vec{r}^{(1)} ] & = \sum_{a=1}^{\nat} m_a \big[
	(\vec{r}_a^{(N)} \! \cdot \eckR_a^{(1)}) \delta_{ij} - 
	r_{a,i}^{(N)}  \widetilde{r}_{a,j}^{(1)} \big],
\end{align}
which is symmetric because of the rotational Eckart conditions.
This finally allows us to approximate the integrand in 
\eqn{eq:Omega-int} by a 
Gaussian, so that
\begin{align}
	& \!\!\!\! \int \! \rmd \bm{\Omega} \, \exp \mleft[-\frac{\beta_N \omega_N^2}{2} 
	\sum_{\mathclap{a=1}}^{\nat} m_a \mleft \lVert 
	\vec{r}^{(N)}_{a} \! - \op{R}(\bm{\Omega})\vec{r}^{(1)}_{a} 
	\mright\rVert^2 \mright] \sim {}  \\
	& \mkern-10mu\mleft(
	\frac{2 \pi}{\beta_N \omega_N^2}
	\mright)^{\mathrlap{\!3/2}} \frac{1}{\det \mat{\eckmom}^{1/2}} 
	\exp\mleft[
	-\frac{\beta_N \omega_N^2}{2} \sum_{a=1}^{n} m_a \mleft\lVert 
	\vec{r}_a^{(N)} - \eckR_a^{(1)} \mright\rVert^2
	\mright]. \nonumber
\end{align}
Under this steepest-descent approximation (which becomes exact as $N \to \infty$), 
\begin{subequations}
\label{eq:eck-partf}
\begin{align}
	Z^{J=0} & = (2 \pi \hbar)^{-N f} \mkern-6mu  \int \rmd \cvec{p} 
	\int \rmd 
	\cvec{r} \, u(\bm{r}) \mkern2mu
	\eu{-\beta_N 
	\rotprojH_{N} (\cvec{p}, \cvec{r})}, \\
	\rotprojH_{N}  & = \funOpN[\text{(open)}]{H} + 
	\sum_{a=1}^{n} \frac{m_a \omega_N^2}{2} \mleft\lVert 
	\vec{r}_a^{(N)} - \eckR_a^{(1)} \mright\rVert^2,
	\label{eq:eck-spring} \\
\label{eq:u-prefactor}
u(\bm{r}) & =  \frac{1}{8 \pi^2} \mleft(
		\frac{2 \pi}{\beta_N \omega_N^2}
		\mright)^{\!3/2} \mkern-14mu \frac{1}{\det 
		\mat{\eckmom}[\vec{r}^{(N)},\vec{r}^{(1)}]^{1/2}},
\end{align}
\end{subequations}
giving the final form of the rotationally projected partition function.

The projection introduces into the integrand a prefactor $u$,
proportional to $\det \mat{\eckmom}^{-1/2} $. The matrix 
$\mat{\eckmom}$ is 
approximately equal to the 
molecule's inertia tensor, and its determinant will fluctuate with a relatively
small amplitude over the course of a simulation. The thermal
expectation value of $u$ can, therefore, be 
calculated efficiently using standard PIMD techniques.

The other change is the replacement of the harmonic spring between replica 1 and replica
$N$ with an ``Eckart spring,'' as defined  by the sum in \eqn{eq:eck-spring}. This 
modified spring does not specifically favor $\vec{r}_1$ and $\vec{r}_N$ 
that are close
to each other. Instead, low energies are achieved for replica pairs whose CoMs 
and \emph{shapes} are near coincident, regardless of relative 
orientation. The two end beads will tend to explore the same potential
energy well, but at a range of relative orientations. This stems from the rotational 
averaging that lead us to $Z^{J = 0}$ in the first place. The forces due to the Eckart spring, required to simulate such trajectories, can be readily obtained from the
$\mat{C}$-matrix and the normalized eigenvector $\vec{q}^{(0)}$ associated with
its smallest eigenvalue:
\begin{equation}
	\pder{}{\vec{r}_{b}^{(i)}} \sum_{a=1}^{\nat} m_{a} \mleft\lVert 
	\vec{r}_{a}^{(N)} - \eckR_{a}^{(1)} \mright\rVert^2 = 
	\sum_{\mathclap{\mu,\nu=0}}^{3} \mkern2mu q^{(0)}_\mu 
	\pder{C_{\mu\nu}}{\vec{r}_{b}^{(i)}} 
	q^{(0)}_\nu.
\end{equation}

The same line of reasoning can be applied to the symmetrized partition 
function in \eqn{eq:rot-partf-aJ}. Going through the derivation we find that
\eqn{eq:eck-partf} still applies, but with a Hamiltonian
\begin{equation}
	\label{eq:HN-J0}
	\funOpN[P]{\rotprojH} = \funOpN[\text{(open)}]{\rotprojH} + 
	\sum_{a=1}^{n} \frac{m_a \omega_N^2}{2} \mleft\lVert 
	\op{P}^{-1}\vec{r}_a^{(N)} - \eckR_a^{(1)} \mright\rVert^2,
\end{equation}
where $\op{P}^{-1}$ is the inverse of $\op{P}$. The prefactor is 
similarly modified, now depending on $\vec{\eckmom}[	
\op{P}^{-1}\vec{r}^{(N)}\!, \vec{r}^{(1)}]$ defined as in
\eqn{eq:mat-A}. The inclusion of an Eckart spring with a 
permutation--inversion
operation similarly allows the end beads to assume a range of
relative orientations, except now
they tend to explore \emph{different} potential energy wells,
with minima related by the operation~$\op{P}$.

\subsection{Thermodynamic Integration%
\label{sec:ti}}

The general expression for the tunneling splitting, \eqn{eq:Delta-mn}, 
and 
its particular realizations for the double [\eqn{eq:dwell-Delta}] and 
triple well
[\eqn{eq:C3-Delta}] are (or can be) written in terms of ratios of 
symmetrized partition 
functions. These functions are of the form
\begin{equation}
	Z_{P}^{J=0} = (2 \pi \hbar)^{-N f} \int \! \rmd 
	\cvec{p}
	\int \! \rmd \cvec{r} \ 
	u (\cvec{r})\, \eu{
		-\beta_N \funOpN[P]{\rotprojH}(\cvec{p}, \cvec{r})
	}.
\end{equation}
The ratio of two partition functions that are symmetrized by the
operators $\op{P}_1$ and $\op{P}_0$, respectively, can be expressed as $Z^{\smash{J=0}}_{P_1} / 
Z^{\smash{J=0}}_{P_0} = A \mkern2mu  \eu{\smash{-\beta_N \Delta F}}$, 
where
\begin{gather}
	\label{eq:partf-ratio}
	\eu{-\beta_N \Delta F} = \frac{
		\displaystyle\int \! \rmd \cvec{p} \int \! \rmd \cvec{r} \ 
		 \eu{ -\beta_N \funOpN[P_1]{\rotprojH}(\cvec{p}, 
		 \cvec{r}) }
	}{
		\displaystyle\int \! \rmd \cvec{p} \int \! \rmd \cvec{r} \ 
		\eu{ -\beta_N \funOpN[P_0]{\rotprojH}(\cvec{p}, \cvec{r}) 
		}
	}
\end{gather}
and $A = \Therm{u_1(\cvec{r})}{}_{P_1} / 
\Therm{u_0(\cvec{r})}{}_{P_0}$. Here, $u_i(\cvec{r})$
refers to \eqn{eq:u-prefactor} evaluated with 
$\vec{\eckmom}[\op{P}_{i}^{\smash{-1}}
\vec{r}^{\smash{(N)}}\!,
\vec{r}^{\smash{(1)}}]$,
and
\begin{gather}
	\Therm[P]{u(\cvec{r})} = 
	\frac{
		\displaystyle\int \! \rmd \cvec{p} \int \! \rmd \cvec{r} \ 
		u(\cvec{r}) \, \eu{ -\beta_N 
		\funOpN[P]{\rotprojH}(\cvec{p}, \cvec{r}) }
	}{
		\displaystyle\int \! \rmd \cvec{p} \int \! \rmd \cvec{r} \ 
		\eu{ -\beta_N \funOpN[P]{\rotprojH}(\cvec{p}, \cvec{r}) }
	} \label{eq:therm-avg}
\end{gather}
denotes the thermal average of $u$ for a system described by the 
Hamiltonian $\funOpN[P]{\rotprojH}$. Such 
averages
are directly available from standard PIMD simulations. The term in
\eqn{eq:partf-ratio} presents more of a challenge;
it is a ratio of partition functions, which we get by
calculating the associated free energy difference, $\Delta F$, using 
thermodynamic integration (TI)\@. To this end
we define
\begin{subequations}
\begin{align}
	Z_{\xi} & = (2 \pi \hbar)^{-Nf} \int \! \rmd \cvec{p} \int \! \rmd \cvec{r} \
	\eu{-\beta_N \rotprojH_{\xi}} \\
	\rotprojH_{\xi} & = (1 - \xi)  \funOpN[P_0]{\rotprojH} + 
	\xi \funOpN[P_1]{\rotprojH}\label{eq:ham-xi}
\end{align}
\end{subequations}
and use the relation
\begin{align}
	\Delta F & = -\frac{1}{\beta_N} \int_0^1 \! \der{\ln Z_{\xi}}{\xi} 
	\, \rmd \xi
	= \int_0^1 \!
	\Therm[\!\xi]{ \funOpN[P_1]{\rotprojH} - 
	 \funOpN[P_0]{\rotprojH}} \, \rmd \xi. \label{eq:ti}
\end{align}
In practice, the integral over $\xi$ is approximated using a numerical quadrature
scheme, which prescribes a set of grid points at which to 
calculate the integrand. As in previous work,\cite{
Matyus2016tunnel1,Zhu2022trimer,Vaillant2019water,Vaillant2018dimer}
 we find that Clenshaw--Curtis (CC) and Gauss--Legendre (GL)
 quadrature\cite{NumRep} both 
offer a substantial improvement over simpler schemes, e.g., Simpson's 
rule.
At every grid point, the integrand is in the form of a thermal average 
for 
a system described by the mixed Hamiltonian $\rotprojH_{\xi}$. Most of 
the terms in the
average cancel exactly: the only remaining contribution
is the difference of potential energies in 
$ \funOpN[\smash{P_1}]{\rotprojH}$ 
and $ \funOpN[\smash{P_0}]{\rotprojH}$ due to the springs 
connecting replica~1 and replica~$N$.

Having obtained all the necessary averages, we calculate their weighted sum with
the weights prescribed by the quadrature scheme. This yields a 
numerical
approximation to the $\xi$-integral in \eqn{eq:ti} and, hence, a value for
$\Delta F$. PIMD simulations at $\xi = 0 $ and $1$
produce the prefactor $A$, which combines with $\Delta F$ 
to give us the target ratio $Z^{\smash{J=0}}_{P_1} / 
Z^{\smash{J=0}}_{P_0} = A \mkern2mu  \eu{\smash{-\beta_N \Delta F}}$. 
This can be substituted
into \eqn{eq:Delta-mn}, to give the desired tunneling splitting 
constant.

%%% FOOTNOTE - used in paragraph below
\footnotetext[25]{Some of the previous tunneling splitting PIMD 
calculations used a normal-mode propagator\cite{Matyus2016tunnel1}
and the more efficient path-integral Langevin equation thermostat.\cite{Vaillant2018dimer} The Eckart spring prevents us from 
directly employing these in our calculations,
but suitable analogous algorithms can probably be developed.
}
%%%

The entire process consists in running a set of PIMD simulations
(no wavefunction calculations involved). To propagate the dynamics and ensure proper
sampling, we use a multiple-time-step (MTS) integrator combined with an Andersen thermostat.\cite{TuckermanBook,Note25} 
The ``slow'' force, used to update the momenta at time 
intervals
of $\Delta t$, derives from the ``external'' potential
$\sum_i V(\vec{r}_i)$. The ``fast'' force, used to update momenta
at intervals of $\delta t$, derives from the
spring potential, which includes
the Eckart spring and a permutation of atomic indices.
The procedure used to obtain confidence intervals for the
reported tunneling splittings is
described in \app{error}.

\section{Results%
\label{sec:results}}

\subsection{Water
\label{sec:water}}

An isolated water molecule (\ce{H2O}) does not exhibit a tunneling 
splitting. However, we can still use this
system as a test of our procedure for projecting onto states of zero angular momentum
($J = 0$), as described in \sref{eck-spring}. A water molecule is small enough that its 
quantum energy levels can be calculated numerically using only modest 
computational resources. To this end, we used the DVR3D program suite 
developed by Tennyson and co-workers,\cite{TennysonDVR3D} from which we 
obtained the ratio $Z^{J=0}/Z$ at two different temperatures. As it was 
not our goal to reproduce experimental data, we used the simple
qTIP4P/F potential to model interatomic interactions, with the same parameters as 
in Ref.~\onlinecite{Habershon2009qtip4pf} and with atomic masses set to 
$\SI{1.007825}{\dalton}$ and
$\SI{15.994915}{\dalton}$ for \ce{H} and \ce{O} respectively.

\begin{table}[b]
\caption{\label{tab:water}%
$Z^{J=0}/Z \times 10^2$ for an isolated molecule of qTIP4P/F water. 
The number of
replicas used in PIMD calculations is indicated in curly brackets, and 
the TI grid sizes are given by $n_{\xi}$. Errors in the last digit
(in parentheses) are reported at $2\sigma$. The exact
results were obtained by running DVR3D\cite{TennysonDVR3D} and 
summing the resulting Boltzmann factors.%
}
\renewcommand{\arraystretch}{1.75}
\begin{ruledtabular}
	\begin{tabular}{r<{\hspace*{0.5ex}}*{4}{c}{c}}
		\multirow{2}{*}{$T \ [\si{K}]$} & 
		\multicolumn{2}{c}{PIMD\{$N=48$\}} & 
		\multicolumn{2}{c}{PIMD\{$N=64$\}} & \multirow{2}{*}{exact} 
		\tabularnewline
		\cmidrule(lr){2-3} \cmidrule(lr){4-5}
		  & $n_{\xi} = 17$ & $n_{\xi} = 33$ & $n_{\xi} = 17$ & $n_{\xi} = 33$ &  \tabularnewline\midrule
  		 300 & 1.21(2) & 1.21(1) & 1.20(1)  & 1.21(1)  & 1.204 \tabularnewline
		 200 & 2.20(2) & 2.20(1) & 2.20(3) & 2.20(2) & 2.200 \tabularnewline
	\end{tabular}
\end{ruledtabular}
\end{table}

PIMD simulations were all performed with a conservative integration 
time step of $\SI{0.1}{fs}$ and an MTS factor $\Delta t / \delta t = 
8$. A CC quadrature grid\cite{NumRep} with $n_{\xi} 
= 33$ points was used for thermodynamic integration. This scheme is 
particularly convenient for running convergence tests,
since a 17-point CC grid can 
be constructed by taking every other point in a 33-point grid.
In \Eqn{eq:partf-ratio}, we replaced
$ \funOpN[\smash{P_0}]{H}$ with $\funOpN{H}$, the standard ring-polymer 
Hamiltonian. The corresponding prefactor 
$u_0(\cvec{r})$ is 1. For $ \funOpN[\smash{P_1}]{\rotprojH} $ we used 
$\funOpN{\rotprojH}$ from \eqn{eq:eck-spring}, and $u_1$ was defined
as in \eqn{eq:u-prefactor}. Every set of TI 
calculations consisted of 32
independent trajectories initialized at the equilibrium geometry
and propagated in parallel. Each one was thermalized
for $t_{\mathrm{th}} = \SI{5}{ps}$ under $H_N$, subject to an Andersen 
thermostat with an average resampling time of $\SI{25}{fs}$ for every 
particle. 
The same propagator and thermostat were used for production 
simulations, which ran sequentially
over $H_{\xi}$ [\eqn{eq:ham-xi}] in order of increasing $\xi$,
starting at $\xi = 0$. At every 
new grid point,
we ran a trajectory of $t_{\mathrm{run}} = \SI{25}{ps}$, starting from 
the configuration at which the simulation for the preceding grid point 
had completed. 
The energy differences in \eqn{eq:ti} were recorded at intervals of $\SI{2}{fs}$, and
values of the prefactor in \eqn{eq:u-prefactor} were recorded every 
$\SI{5}{fs}$. Data
from the first $\SI{500}{fs}$ of every production trajectory were discarded.

Our simulation results are summarized in \tref{water}, from which we 
can see 
that $n_{\xi} = 17$ CC quadrature points and $N = 48$ replicas are sufficient 
to converge the PIMD results, with an estimated statistical error $< 1\%$. 
At both simulation temperatures, the results are within one standard 
deviation
from the exact (DVR3D) values, verifying our procedure for projecting 
onto $J = 0$ states.

\subsection{Hydronium%
\label{sec:h3o}}

The hydronium ion (\ce{H3O+}) is one of the simplest molecules to 
exhibit tunneling splitting, similar to the isoelectronic 
\ce{NH3} but with a lower barrier and, correspondingly, 
larger splitting. The inversion barrier is sufficiently low 
and anharmonic for instanton theory 
to be in error by over 20\%,\cite{Cvitas2018instanton}
making hydronium a suitable test system for our method.
To describe the interatomic interactions, we have chosen
the PES developed by 
{\NoHyper\citeauthor{Bowman2016Hydronium}\endNoHyper},\cite{Bowman2016Hydronium}
 which was fitted to a set of \mbox{CCSD(T)-F12b/aug-cc-pVQZ} 
 electronic 
energies and has a barrier of \SI{674.3}{cm^{-1}}.
The lowest vibrational frequency of approximately \SI{887}{cm^{-1}}
corresponds to the tunneling umbrella mode, from which it can be deduced
that only one pair of tunneling-split $J=0$ states reside below the 
barrier.\footnote{In fact, the lower of the two states is forbidden
by nuclear spin symmetry.\cite{BunkerBook} Of course, we can still
calculate the corresponding tunneling
splitting and compare to the result in
Ref.~\citenum{Verhoeve1989H3Osplit}, obtained by fitting a 
rovibrational Hamiltonian to experimental data.}

\begin{figure}[b]
	\includegraphics{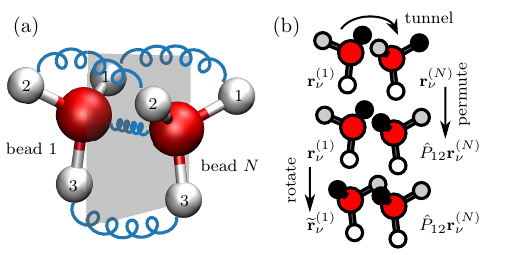}
	\caption{%
		(a)~A low-energy hydronium (\ce{H3O+}) ``polymer'' subject 
		to an 
		Eckart spring
		with a $(12)$ permutation [final term in \eqn{eq:HN-J0}].
		The hydrogen atoms of beads $1$ and $N$ are labeled with the 
		corresponding indices, and intermediate beads, spread out for 
		clarity,
		are represented with coiled blue lines; in reality, the oxygen 
		atoms
		are nearly coincident. A gray rectangle shows the mirror plane 
		that 
		approximately maps the end beads onto each other. This operation
		is equivalent to the permutation of atoms and rotation
		shown on the right.
		(b)~A schematic illustrating the calculation of the Eckart spring 
		potential.
		In the starting configuration (top), the atoms of the $N$th 
		replica are 
		permuted (middle), and the 1st replica is rotated about 
		its center of mass (bottom), to match the permuted replica as 
		closely 
		as possible. A good match implies a low energy of the Eckart 
		spring.%
		\label{fig:h3o-eckart}}
\end{figure}

The molecular symmetry group of hydronium is isomorphic to 
$D_{3h}$,\cite{Schnell2011Groups} and its characters can 
be directly substituted into \eqn{eq:Delta-mn} 
to produce an expression for the tunneling splitting. However,
as noted in the discussion after \eqn{eq:Delta-mn}, 
the analysis can also be performed using the 
smallest subgroup that distinguishes between the tunneling-split
energy levels. In the present case we use the $\{E, \, (12) \}$ 
subgroup (isomorphic to $C_s$), whose (12) operation permutes hydrogens~1 and~2 (see \fig{h3o-eckart} for the atomic indices). The 
general 
expression for the tunneling splitting then reduces to 
the double-well formula in \eqn{eq:dwell-Delta}, where
$\op{\sigma}$ is replaced with $\op{P}_{12}$, the $(12)$ permutation. 
TI was 
performed using 
$\funOpN{\rotprojH}$ [\eqn{eq:eck-spring}] and 
$\funOpN[\smash{(12)}]{\rotprojH}$ 
[\eqn{eq:HN-J0} with $\op{P}^{\smash{-1}} = \op{P}_{12}$] for 
the $\funOpN[\smash{P_0}]{H}$ and $\funOpN[\smash{P_1}]{H}$ 
Hamiltonians, respectively.
The corresponding prefactor functions were
$ u_0 = \det \mat{\eckmom}[\vec{r}^{\smash{(N)}}\! , 
\vec{r}^{\smash{(1)}}]^{\smash{-1/2}} $ 
and 
$ u_1 = \det \mat{\eckmom}[\op{P}_{12} \vec{r}^{(N)}\! , 
\vec{r}^{(1)}]^{-1/2} $. 
The same PIMD simulation parameters as in \sref{water} were used for 
this section, except that the average resampling time in the Andersen 
thermostat was increased to \SI{100}{fs} and Gauss--Legendre (GL) 
quadrature\cite{NumRep} was used instead of CC, as it was found
to converge more rapidly with grid size.

For \eqn{eq:dwell-Delta} to give an accurate estimate of the tunneling
splitting, the dynamics have to be simulated at a sufficiently low 
temperature. The lowest vibrational frequency of \SI{887}{cm^{-1}} 
corresponds to a temperature of \SI{1280}{K}. At this temperature
we can expect some excited vibrational states to have a significant 
thermal population. Hence, we reduced this temperature by 
a factor of approximately six, taking \SI{200}{K} as our 
starting point. We then ran PIMD simulations for a series of 
progressively lower 
temperatures, taking care to converge the estimated tunneling 
splittings with respect to the number of beads, $N$, and the 
number of GL grid points, $n_{\xi}$, at each temperature
(see \tref{h3o}). Convergence
is already satisfactory at \SI{150}{K} for $N=96$ and $n_{\xi} = 15$,
and the final value of $\Delta = \SI{53.2(4)}{cm^{-1}}$ is taken from
the simulation at $T = \SI{100}{K}$,  $N = 128$, and $n_{\xi} = 20$
(since $\Delta$ was judged converged at this grid size and
the simulation with $n_{\xi} = 30$ involved fewer, shorter trajectories 
and, therefore, resulted in a larger statistical error).

\begin{table}[t]
	\caption{\label{tab:h3o}%
		Ground-state tunneling splittings calculated with PIMD for
		the hydronium potential from Ref.~\citenum{Bowman2016Hydronium}.
		$T$ specifies the temperature, $N$ gives the number 
		of beads,
		and $n_{\xi}$ the number of Gauss--Legendre quadrature points.
		Errors in the last digit (in parentheses) are reported at 
		$2\sigma$.
		In the bottom part of the table, our most reliable
		result (PIMD; $T = \SI{100}{K}$,  $N = 128$, $n_{\xi} = 20$ --- 
		see main text) is  compared  against instanton (inst.), 
		vibrational configuration interaction 
		(VCI)
		and experimentally measured (expt.) tunneling splittings.%
	}
	\renewcommand{\arraystretch}{1.50}
	\begin{ruledtabular}
		\begin{tabular}{*{5}{c}}
			\multirow{2}{*}{$T \ [\si{K}]$} &
			\multirow{2}{*}{$N$} & 
			\multicolumn{3}{c}{$\Delta \ [\SI{}{\per\cm}]$} 
			\tabularnewline
			\cmidrule(lr){3-5} 
			& & $n_{\xi} = 15$ & $n_{\xi} = 20$ & $n_{\xi} = 30$ 
			\tabularnewline\midrule
			\multirow{2}{*}{200} &
			48 & 55.2(7) & 55.7(6) & 55.2(5) \tabularnewline
			&
			64 & 54.5(8) & 54.6(8) & 54.2(6) \tabularnewline
			\multirow{2}{*}{150\footnotemark[1]} &
			64 & 53.7(3) & 53.8(3) & 53.4(5) \tabularnewline
			&
			96 & 53.1(3) & 52.9(3) & 52.8(6) \tabularnewline
			\multirow{2}{*}{100\footnotemark[1]} &
			96 & 53.3(4) & 53.7(4) & 53.3(6) \tabularnewline
			&
			128 & 53.2(4) & 53.2(4) & 53.2(7) \tabularnewline
			\midrule
			method &  \textbf{inst.}\footnotemark[2] &  
			\textbf{VCI}\cite{Bowman2016Hydronium} & 
			\textbf{PIMD}\footnotemark[2] & 
			\textbf{expt.}\cite{Verhoeve1989H3Osplit}  \tabularnewline
			$\Delta \ [\SI{}{\per\cm}]$ & 69.8 & 52.48 & 53.2(4) &  
			55.35 
			\tabularnewline
		\end{tabular}
	\end{ruledtabular}
	\footnotetext[1]{For $n_{\xi} = 15$ and $20$, the number of 
	independent 
		trajectories was increased to 64 and their duration doubled to 
		\SI{50}{ps} per quadrature point.}
	\footnotetext[2]{This work.}
\end{table}

%%% FOOTNOTE - used in paragraph below
\footnotetext[26]{The calculation is sensitive to the accuracy of
the potential energy gradient and Hessian. To achieve satisfactory
numerical stability, fourth-order central differences\cite{NumRep} were 
used 
to calculate first and second derivatives.
}
%%%

To put this value into context, we compare it against the instanton
tunneling splitting, which we calculated following the protocol in 
Ref.~\citenum{InstReview}.\cite{Note26} The calculation converges to 
\SI{69.8}{cm^{-1}} at $\beta = \SI{8000}{\hartree^{-1}}$ ($T \approx 
\SI{40}{K} $) for $N = 
1024$, overestimating the PIMD value by over 30\%.\footnote{%
Note that the instanton calculation requires a lower temperature
and larger number of beads to reach convergence. This
is necessary to satisfy the assumptions involved in integrating
over the permutational mode and neglecting the tunneling time
parameter\cite{RPInst,Matyus2016tunnel1,InstReview,tunnel}
}
Our other comparison
is to the vibrational configuration interaction (VCI) value of
\SI{52.48}{cm^{-1}} calculated
by 
\begin{NoHyper}\citeauthor{Bowman2016Hydronium}%
\end{NoHyper}.\cite{Bowman2016Hydronium}
There is a small but statistically significant difference between that 
and our PIMD result, which may indicate that VCI is not sufficiently
converged with basis set size or with respect to the number of 
modes used to represent the vibrational angular momentum terms.
Even so, this discrepancy
is small compared to the deviation from the experimentally measured 
splitting of \SI{55.35}{cm^{-1}}, which both
approaches underestimate  by \SIrange{2}{3}{cm^{-1}}. It is likely
that the deviation is caused by imperfections in the PES, and
we provide evidence in the supplementary material that it arises from a 
combination of fitting and basis-set truncation errors.

\subsection{Methanol%
\label{sec:h3coh}}

\begin{figure}[b]
	\includegraphics{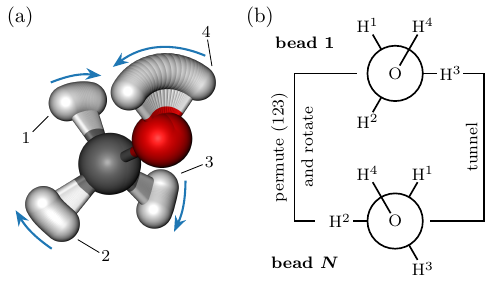}
	\caption{%
		(a)~A minimum-energy methanol (\ce{CH3OH}) ``polymer''
		subject to an Eckart spring with a $(123)$ permutation.
		The hydrogen atoms of the first replica are labeled with the 
		corresponding indices; taken together, the beads trace
		out the torsional tunneling pathway, and the direction of 
		motion is indicated with blue arrows.
		(b)~Newman projections of the first
		and final replicas (beads). Here, the replicas
		are mapped onto each other exactly
		by a $(123)$ permutation and rotation about the center of mass
		and, therefore, the Eckart spring is fully contracted
		[see also \fig{h3o-eckart}(b)].
		\label{fig:meoh}}
\end{figure}

Methanol (\ce{CH3OH}) is a multi-well system with a MS group that is isomorphic to 
$C_{3v}$ (see \sref{multi-well} and \fig{c3v-pot}).
Its ground state is split into an $A_1$ and a doubly degenerate
$E$ level by torsional tunneling through 
a relatively low potential energy barrier.\footnote{%
Both states are allowed by nuclear spin symmetry and have
equal spin statistical weights\cite{BunkerBook}}
With six atoms, methanol
is large enough to present a challenge to wavefunction-based
approaches,\cite{Lauvergnat2014Methanol,Carrington2017quantum}
and sufficiently anharmonic to introduce a sizable error into
the instanton approximation.

To run our PIMD simulations of methanol we turned to the
potential by 
\begin{NoHyper}\citeauthor{Qu2013Methanol}\end{NoHyper},\cite{Qu2013Methanol}
fitted to \mbox{CCSD(T)-F12b/aug-cc-pVDZ} energies for configurations
relevant to the present study, with
a torsional barrier of \SI{350}{cm^{-1}} and a harmonic torsional
frequency of \SI{286}{cm^{-1}}. Following \sref{multi-well},
symmetry analysis was conducted in the $\{ E,\, (123), \,(132)\}$
subgroup, which is isomorphic to the $C_3$ group. The tunneling
splitting can, therefore,
be calculated using \eqn{eq:C3-Delta}, where it is 
understood that all constituent partition functions are projected onto
$J = 0$ and the $C_3$ operation is replaced by the $(123)$ cyclic 
permutation. Our indexing convention is shown in \fig{meoh}, 
along with the dominant polymer configuration, corresponding
to the minimum of $\funOpN[\smash{(132)}]{\rotprojH}$ 
[\eqn{eq:HN-J0} with $\op{P}^{\smash{-1}} = \op{P}_{123}$]. The 
tunneling
pathway traced out by the polymer beads highlights the large extent
to which the \ce{OH} hydrogen participates in the torsional 
motion of the \ce{CH3} group.

\begin{table}[t]
	\caption{\label{tab:MeOH}%
		Ground-state tunneling splittings calculated with PIMD for the
		methanol potential from Ref.~\citenum{Qu2013Methanol}. Column
		headings and error specification follow the convention
		used in \tref{h3o}. In the bottom part of the table,
		the final PIMD result is compared against instanton (inst.), 
		wavefunction (WF) and experimental (expt.) tunneling splittings.
	}
	\renewcommand{\arraystretch}{1.50}
	\begin{ruledtabular}
		\begin{tabular}{cd{1}*{3}{d{6}}}
			\multirow{2}{*}{$T \ [\si{K}]$} &
			\multirow{2}{*}{$N$} & 
			\multicolumn{3}{c}{$\Delta \ [\SI{}{\per\cm}]$} 
			\tabularnewline
			\cmidrule(lr){3-5} 
			& & 
			\multicolumn{1}{c}{$n_{\xi} = 5$} &
			\multicolumn{1}{c}{$n_{\xi} = 10$} &
			\multicolumn{1}{c}{$n_{\xi} = 15$}
			\tabularnewline\midrule
			\multirow{2}{*}{75\footnotemark[1]} &
			64 & 8.4(2) & 8.39(18) & \multicolumn{1}{c}{---} 
			\tabularnewline
			&
			96 & 8.2(3) & 8.28(14) & 8.33(11) \tabularnewline
			\multirow{2}{*}{50\footnotemark[2]} &
			96 & 9.18(15) & 9.15(13) & 9.21(8) \tabularnewline
			&
			128 & 9.08(12) & 9.05(11) & 9.03(8) \tabularnewline
			\multirow{2}{*}{35\footnotemark[2]} &
			128 & 9.2(2)  & 9.10(13) & 9.09(11) \tabularnewline
			&
			196 & 9.00(19) & 9.1(3) & 9.14(13)  \tabularnewline
			\midrule
			\multicolumn{1}{c}{method} & 
			\multicolumn{1}{c}{\textbf{inst.}\footnotemark[4]} &  
			\multicolumn{1}{c}{\textbf{WF}\footnotemark[3]} & 
			\multicolumn{1}{c}{\textbf{PIMD}\footnotemark[4]} &   
			\multicolumn{1}{c}{\textbf{expt.}\cite{Xu1997MeOHsplit}}  
			\tabularnewline
			\multicolumn{1}{c}{$\Delta \ [\SI{}{\per\cm}]$} &
			\multicolumn{1}{c}{11.3} &
			\multicolumn{1}{c}{9.15} &
			\multicolumn{1}{c}{9.14(13)} & 
			\multicolumn{1}{c}{9.12}
			\tabularnewline
		\end{tabular}
	\end{ruledtabular}
	\footnotetext[1]{Thermalized for \SI{5}{ps}, propagated for 
		\SI{25}{fs} 
		at every grid point, 32 independent trajectories.}
	\footnotetext[2]{Thermalized for \SI{10}{ps}, propagated for 
		\SI{50}{fs} at every grid point, 64 independent trajectories.}
	\footnotetext[3]{Value by 
		\begin{NoHyper}\citeauthor{Lauvergnat2014Methanol}\end{NoHyper}%
		\cite{Lauvergnat2014Methanol} obtained from
		sparse-grid wavefunction
		calculations on the PES from 
		Ref.~\citenum{Bowman2007methanol}.}
	\footnotetext[4]{This work.}
\end{table}

Thermodynamic integration for this system was performed between
$\funOpN{\rotprojH}$ and $\funOpN[(132)]{\rotprojH}$, with prefactor
functions $ u_0 = \det \mat{\eckmom}[\vec{r}^{(N)}\! , 
\vec{r}^{(1)}]^{-1/2} $ and 
$ u_1 = \det \mat{\eckmom}[\op{P}_{123} \vec{r}^{(N)}\! , 
\vec{r}^{(1)}]^{-1/2} $, using a GL quadrature grid. We used a 
simulation
setup analogous to \sref{h3o}, with a propagation time step of 
\SI{0.2}{fs},
an MTS factor of $8$, and a mean resampling time of \SI{100}{fs}.
Further simulation details and calculated tunneling splitting values
are given in \tref{MeOH}.

Given the lowest vibrational frequency of 
\SI{286}{\cm^{-1}} (\SI{411}{K}), we started our 
simulations at \SI{75}{K} and progressively lowered the temperature
until reaching convergence. Our most reliable result was calculated
at a temperature of \SI{35}{K} with $N = 196$, giving a final value
of $\Delta = \SI{9.14(13)}{cm^{-1}}$.
The corresponding instanton tunneling 
splitting calculation\cite{InstReview,Note26} converged to 
\SI{11.3}{cm^{-1}} at $\beta = \SI{20000}{\hartree^{-1}}$ ($T \approx 
\SI{16}{K}$) and $N = 1024$. 

We are not aware of any wavefunction-based
tunneling splitting calculations for the PES employed in this work. 
However, an earlier PES by Bowman and 
co-workers,\cite{Bowman2007methanol} fitted to CCSD(T)/aug-cc-pVTZ
energies at many of the same molecular configurations,
is believed to achieve similar accuracy. This PES was used
by \begin{NoHyper}\citeauthor{Lauvergnat2014Methanol}\end{NoHyper}
in Ref.~\citenum{Lauvergnat2014Methanol} to calculate the splitting
by solving the Schr\"{odinger equation on a sparse grid}. Our PIMD estimate is in excellent
agreement with their result of \SI{9.15}{cm^{-1}}, and
both are very close to the experimentally measured 
splitting\cite{Xu1997MeOHsplit} of \SI{9.12}{cm^{-1}}.
Compared to the results in \sref{h3o}, this level of agreement
is surprising given the relatively small basis set used to calculate
the underlying ab~initio energies. It seems that the PES benefits
from a fortuitous cancellation between fitting and basis-set
truncation errors, as shown in the supplementary material.
Even so, the key point is that the PIMD method has achieved
its goal of probing the accuracy of the PES.

\section{Conclusions%
\label{sec:conclusions}}

We have presented an exact path-integral method for calculating 
ground-state rovibrational tunneling splittings from ratios of 
symmetrized partition functions. By construction, splitting values
are the only unknown parameters in our approach, and a rigorous
projection onto the $J = 0$ state is achieved by means of an Eckart 
spring (avoiding the need to integrate over orientations explicitly). 
The first feature makes it possible to extract tunneling splittings
from just a single thermodynamic integration, offering 
a simplification over the procedure used by M\'atyus 
and co-workers.\cite{Matyus2016tunnel1,Matyus2016tunnel2}
The second feature means that rotational projection is
performed in an exact and computationally efficient manner, 
unlike in previous PIMD studies,\cite{
Matyus2016tunnel1,Vaillant2018dimer,Vaillant2019water,Zhu2022trimer}
such that we can fully account for rovibrational coupling effects.

A perturbatively corrected instanton (RPI+PC)
calculation\cite{Lawrence2023PC} would likely approach the 
accuracy of our PIMD method for the molecular systems
in this study. However, whereas the perturbative correction
will eventually fail, the PIMD calculation should remain robust even
in extremely low-barrier systems, for which large-amplitude motions
correspond to above-barrier 
states.\cite{Sarka2016H5interpret,Wang2015CH3NO2}
Additionally, unlike the current formulation of RPI+PC, our method
can be readily extended to $J > 0$ states, so that in the future
we hope to use Eckart-spring PIMD to study the dependence of 
tunneling splitting on the rotational state.\cite{Verhoeve1989H3Osplit}
We should also be able to calculate the energy levels of fluxional
systems, where rotational and vibrational motion cannot be 
separated even to a first approximation, as in the case of the
\ce{CH5+} molecular ion.\cite{Wodraszka2015CH5}

As expected, we have observed that tunneling splittings are very 
sensitive to the underlying PES. Using potentials fitted
to high-accuracy ab~initio energies will, therefore, be crucial if
we are to interpret and compare with experimental data in upcoming 
studies. While for some molecules, including the ones discussed here, 
such potentials have already been developed, the availability of 
an accurate PES can, unfortunately, limit the choice
of system. At least in part, this problem can be alleviated with the
help of machine learning (ML).\cite{Bowman2023MLPerspective} In 
contrast to 
wavefunction-based
approaches, which essentially require a global PES, PIMD only
samples configurations around the instanton(s). One can, therefore, 
fit to high-accuracy ab~initio data around the instanton 
path,\cite{GPR,TransferLearning} expanding the
scope of our method. 

Instantons may also help us optimize the PIMD
procedure itself. A major goal is to minimize the sampling time
needed to achieve a specified level of statistical accuracy.
This may be accomplished with the help of more sophisticated
thermostats\cite{Vaillant2018dimer,Ceriotti2011PIGLE}
or higher-order propagator splittings, beyond the Trotter 
product formula.\cite{Suzuki1985Split,Kapil2016PI} Alternatively, 
one could use the system-specific information available from an 
instanton, which, in a sense, describes the 
optimal tunneling pathway. This may help us devise the
optimal quadrature scheme for thermodynamic integration
or otherwise improve the efficiency of our free-energy calculations.
It should also be noted that, unlike PIMD methods utilizing
\emph{ring} polymers,\cite{TuckermanBook,Ceperley1995PathIntegrals}
we have not fully exploited the cyclic symmetry of the trace.
Doing so should lead to better statistical behavior compared to approaches
based on density-matrix elements.\cite{Matyus2016tunnel1}
The development of this idea is left for future work.

\section*{Supplementary Material}

See supplementary material for thermodynamic integration curves,
intermediate values used in calculating the results in Tables~\ref{tab:water}--\ref{tab:MeOH}, and estimates of systematic
errors due to inaccuracies in the PES.%

\section*{Data Availability}

The data that support the findings of this study are available within the article and its supplementary material.%

\section*{Acknowledgments}

The authors thank Joel Bowman, Qi Yu, and Chen Qu for providing the 
potential energy surfaces for hydronium and methanol. They also 
acknowledge financial support from the Swiss National Science 
Foundation through SNSF Project No.\ 207772.

\section*{Authors' Contribution}

\textbf{George Trenins:} formal analysis (equal), investigation (lead), 
methodology (lead), software (lead), supervision (lead), visualization 
(lead), writing -- original draft (lead), writing -- review (equal). 
\textbf{Lars Meuser:} methodology (supporting), validation (equal).
\textbf{Hannah Bertschi:} methodology (supporting), software (supporting).
\textbf{Odysseas Vavourakis:} methodology (supporting), validation (equal), visualization (supporting).
\textbf{Reto Fl\"{u}tsch:} formal analysis (equal), software (supporting), investigation (supporting).
\textbf{Jeremy Richardson:} conceptualization (lead), supervision (supporting), writing -- original draft (supporting), writing -- review (equal).

\appendix

\section{Error propagation
\label{app:error}}

The standard error, $\sigma_{u}$, of the mean of 	
$\Therm[\op{P}]{u(\cvec{r})} $ 
[\eqn{eq:therm-avg}], is
estimated from the variance of the values obtained from 
independent trajectories.  For $\Delta F$ [\eqn{eq:ti}], the 
error $\sigma_{\Delta F}$ is estimated using the 
bootstrap method.\cite{NumRep} These estimates are combined
to get the standard error in the ratio $\Zratio = Z_{P_1}^{\smash{J=0}} 
/
 Z_{P_0}^{\smash{J=0}}$
according to
\begin{align}
\sigma_{\Zratio}  = \Zratio & {} \times 
\mleft\{
\mleft(
\sigma_{u_0} \big / \Therm[\op{P}_0]{u_0(\cvec{r})}
\mright)^{\!2} + {}
 \mright. \\
& {} \qquad \mleft.
\mleft(
\sigma_{u_1} \big / \Therm[\op{P}_1]{u_1(\cvec{r})}
\mright)^{\!2} + 
\mleft(\beta_N \mkern1mu \sigma_{\mkern-1mu \Delta F} \mright)^2
\mright\}^{\!1/2}.
 \nonumber
\end{align}
Then, for a double-well tunneling splitting derived
from \eqn{eq:dwell-Delta} (as in \sref{h3o}), the standard error
is
\begin{equation}
\sigma_{\Delta} = \frac{2\sigma_{\Zratio}}{\beta(1 - \Zratio^2)}.
\end{equation}
For a triple-well splitting derived from \eqn{eq:C3-Delta} 
(as in \sref{h3coh}), the error is
\begin{equation}
\sigma_{\Delta} = \frac{3\sigma_{\Zratio}}{\beta(1 - \Zratio)(1 + 
2\Zratio)}.
\end{equation}

\bibliography{references,extra}

\end{document}